\documentclass[12pt]{iopart}

\usepackage{iopams}
\usepackage{graphicx}

\begin{document}

\title[Lattice-based ground state transfer]{Deeply bound ultracold molecules in an optical lattice}

\author{Johann G. Danzl$^a$, Manfred J. Mark$^a$, Elmar Haller$^a$, Mattias Gustavsson$^a$, Russell Hart$^a$, Andreas Liem$^b$, Holger Zellmer$^c$, and Hanns-Christoph N\"agerl$^a$}

\address{$^a$ Institut f{\"u}r Experimentalphysik und Zentrum f\"{u}r Quantenphysik, \\
Universit{\"a}t Innsbruck, Technikerstra{\ss}e 25, A--6020 Innsbruck, Austria \\
$^b$ JT Optical Engine GmbH + Co. KG,\\
Pr{\"u}ssingstra{\ss}e 41, D--07745 Jena, Germany \\
$^c$ Hochschule f{\"u}r Technik, Wirtschaft und Kultur Leipzig,\\
Gutenbergplatz 2-4, D--04103 Leipzig, Germany}
\ead{Christoph.Naegerl@uibk.ac.at}
\begin{abstract}
We demonstrate efficient transfer of ultracold molecules into a deeply bound rovibrational level of the singlet ground state potential in the presence of an optical lattice. The overall molecule creation efficiency is 25\%, and the transfer efficiency to the  rovibrational level $|v\!=\!73, J\!=\!2\!>$ is above 80\%. We find that the molecules in $|v\!=\!73,J\!=\!2\!\!>$ are trapped in the optical lattice, limited by optical excitation by the lattice light. The molecule trapping time for a lattice depth of 15 atomic recoil energies is about 20 ms. We determine the trapping frequency by the lattice phase and amplitude modulation technique. It will now be possible to transfer the molecules to the rovibrational ground state $|v\!=\!0,J\!=\!0\!\!>$ in the presence of the optical lattice.

\end{abstract}

\maketitle

\section{Introduction}
The generation of molecular quantum gases and molecular Bose-Einstein condensates (BEC) has been a major goal for the field of atomic and molecular physics. It has been achieved for the case of two fermionic atoms that pair up to form a bosonic dimer molecule in the limit of vanishing binding energy \cite{Jochim2003,Greiner2003} at ultralow temperatures. In this limit, collisional stability is assured, and this has allowed the investigation of the BEC-BCS crossover \cite{Fermi2008}. Here, we are interested in the opposite limit of deeply bound molecules. Collisional stability is expected only for the rovibronic ground state, and most likely it will be necessary that one prepares the lowest molecular hyperfine sublevel \cite{Aldegunde2008} to avoid hyperfine changing collisions. Our approach to producing a quantum gas of ground state molecules is based on laser cooling of atoms to the point of quantum degeneracy, followed by molecule association on a Feshbach resonance and subsequent coherent two-photon molecule transfer \cite{Danzl2008a,Mark2008,Danzl2008b,Winkler2007,Ni2008,Lang2008}. In principle, this approach combines high molecular densities and ultralow temperatures with full state selectivity. For optimization of both the initial molecule creation process and the transfer process, the use of a three-dimensional optical lattice has been proposed, as illustrated in Fig. 1 C. In a superfluid-to-Mott-insulator phase transition doubly occupied lattice sites can be favored \cite{Volz2006,Duerr2008}, and collisional relaxation during the transfer can, at least in principle, be fully avoided. It should be possible that one finally creates a molecular BEC by dynamical melting of the lattice after the two-photon transfer  \cite{Jaksch2002}.

In the present work, we report on two-photon transfer into a deeply bound rovibrational level by means of the stimulated Raman adiabatic passage (STIRAP) technique \cite{Bergmann1998,Winkler2007} in the presence of a three-dimensional optical lattice. We extend our previous work of transfering molecules to rovibrational level $|v\!=\!73, J\!=\!2\!>$ of the $^1\Sigma_g^+$ electronic ground state in the quantum gas regime \cite{Danzl2008a} by first using the superfluid-to-Mott-insulator phase transition to efficiently produce pairs of atoms at the wells of the lattice. The pairs are then associated to weakly bound molecules on a Feshbach resonance. Subsequently, the molecules are transferred by magnetic field ramping to the starting state for optical transfer. From there, they are efficiently transferred to the deeply bound rovibrational level $|v\!=\!73, J\!=\!2\!>$ by means of STIRAP. Fig.\,1 A shows the relevant molecular states for the Cs dimer molecule and the transitions involved. We find that the molecules in $|v\!=\!73, J\!=\!2\!>$ are trapped in the lattice with a $1/e$-trapping time of about $20$ ms, limited by scattering of lattice light. We measure the trapping frequency of the molecules in the lattice and find that the polarizability in $|v\!=\!73, J\!=\!2\!>$ is about 30\% that of the Feshbach molecules. It will now be possible that one adds a second STIRAP transfer step the reach the rovibronic ground state $|v\!=\!0, J\!=\!0\!>$, giving full quantum control over the external and internal degrees of freedom for the molecules.

\section{Preparation of Feshbach molecules in the optical lattice}
To produce an ultracold sample of Feshbach molecules trapped at the individual sites of an optical lattice we first produce an atomic BEC with typically $1\times10^5 $ Cs atoms in the lowest hyperfine sublevel $F\!=\!3, \ m_F\!=\!3$ in a crossed optical dipole trap. As usual, $F$ is the atomic angular momentum quantum number, and $m_F$ its projection on the magnetic field axis. For BEC production, we essentially follow the procedure detailed in Ref.\cite{Weber2003}. We set the atomic scattering length to a value of 210 a$_0$, where a$_0$ is Bohr's radius, by tuning the magnetic offset field to 2.1 mT. At this value, three-body losses are minimal \cite{Kraemer2006}. We then drive the superfluid-to-Mott-insulator phase transition \cite{Greiner2002} by exponentially ramping up the power in a three-dimensional optical lattice within about $400$ ms while simultaneously ramping up the harmonic confinement in the dipole trap.  The lattice is generated by three mutually orthogonal, retro-reflected laser beams at a wavelength of $ \lambda= 1064.5$ nm, each with a $1/e$-waist of about $350 \ \mu$m. For the atoms, we achieve a well depth of up to 40 $E_R$, where $E_R=h^2/(2 m \lambda^2) = k_B \! \times \! 64 \, $nK is the atomic photon recoil energy with the mass $m$ of the Cs atom. $h$ is Planck's constant, and $k_B$ is Boltzmann's constant. Throughout the paper we give lattice depths in units of the atomic recoil energy. The lattice light as well as the light for the dipole trap beams is derived from a single-frequency, narrow-band, highly-stable Nd:YAG laser that is amplified to up to 20 W without spectral degradation in a home-built fiber amplifier \cite{Liem2003}. The power in each lattice beam is controlled by an acousto-optical intensity modulator and an intensity stabilization servo. While ramping up the lattice potential, the power in the two dipole trap beams is increased to assure that the central density in the trap is sufficiently high to allow the preferential formation of atom pairs at the central wells of the lattice, but not too high to lead to triply occupied sites. We typically ramp the lattice to a depth of 15 to 25 $E_R$. Typically about 30\% of the atoms reside at doubly occupied lattice sites. We estimate this number from the molecule production efficiency. This value is not optimal yet, as loading from a parabolic potential should give a maximum of 53\% \cite{Duerr2008,Hansis2006}.

We now produce Feshbach molecules on a Feshbach resonance \cite{Regal2003,Herbig2003,Koehler2006} near a magnetic field value of $B=1.98$ mT \cite{Mark2007} in the presence of the optical lattice \cite{Thalhammer2006,Volz2006}. Fig.\,1 B shows the relevant weakly bound Feshbach levels. The resonance at 1.98 mT is quite narrow, but it lies at a conveniently low value of the magnetic field, allowing us to simply lower the magnetic offset field from the BEC production value and ramp over the resonance with a rate of about $0.006$ T/s. The molecules produced are then in level $|g\!>$. These molecules have $g$-wave character, i.e. $\ell \! = \! 4$, where $\ell$ is the quantum number associated with the mechanical rotation of the nuclei \cite{Chin2004}. After association, atoms remaining at singly occupied lattice sites are removed by microwave transfer to $F=4$ and a resonant light pulse. Starting from level $|g\!>$ we have recently identified transitions to deeply bound excited rovibrational levels of the Cs$_2$ mixed $($A$^1\Sigma_u^+ - $b$^3\Pi_u) \ 0_u^+$ excited states \cite{Danzl2008b}. These transitions should allow STIRAP transfer to the target rovibrational level $|v\!=\!73, J\!=\!2\!>$ of the electronic ground state, but for the present work we have decided to use Feshbach level $|s\!>$ as the starting state as in our previous work \cite{Danzl2008a} so that the transfer performances with and without the presence of the lattice can be compared. To reach level $|s\!>$ from level $|g\!>$, we have implemented Feshbach state transfer as realized in Ref.\cite{Mark2007} using a combination of slow and fast magnetic field ramps. In brief, we first transfer the molecules from $|g\!>$ to level $|g_2\!>$ by lowering the magnetic field $B$ sufficiently slowly to a value of 1.22 mT, thereby following the upper branch of an avoided crossing near 1.33 mT as shown in Fig.\,1 B. We then increase $B$ abruptly to a value of 1.67 mT, thereby jumping the two crossings with levels $|g\!>$ and $|l\!>$. The maximum magnetic field rate of change is $\sim 2000$ T/s. We finally follow slowly on the upper branch of the avoided crossing with $|s\!>$ at 1.85 mT, stopping at $B=1.9$ mT. Our procedure allows us to essentially transfer all molecules from $|g\!>$ to $|s\!>$. For molecule detection, we reverse the magnetic field ramps to level $|g\!>$, dissociate the molecules at the Feshbach resonance at $B=1.98$ mT and detect the resulting atoms by standard absorption imaging \cite{Herbig2003}.

For comparison with our data obtained below we first measure the lifetime of the weakly-bound Feshbach molecules in the optical lattice. Typical lifetime measurements for these molecules are shown in Fig.\,2 A-C. In such measurements, we record the number of remaining molecules as a function of hold time in the lattice. The lifetime of the molecules depends strongly on which Feshbach level is used and on the value of the magnetic field $B$. For example, for molecules in level $|g\!>$ at $B\!=\!1.82$ mT the lifetime is $1.8$ s at a lattice depth of 15 $E_R$, while in level $|s\!>$ the lifetime is $0.09$ s at $B\!=\!1.9$ mT and $10$ s at $B\!=\!2.9$ mT for the same lattice depth. We attribute this strong dependence of the lifetime of molecules in $|s\!>$ to the fact that the molecular character changes strongly from being predominantly closed channel dominated to being open channel dominated as the magnetic field is increased \cite{Koehler2006}, reducing wave function overlap with excited molecular levels. We always determine the lifetime for two values of the lattice depth, 15 $E_R$ and 25 $E_R$. In all cases, the lifetime is reduced for higher lattice depth, indicating residual optical excitation by the lattice light. Nevertheless, the long lifetimes reflect the fact that the lattice perfectly shields the molecules from inelastic molecule-molecule collisions, which would otherwise limit the lifetime to a few ms at the given molecular densities \cite{Thalhammer2006}.

\section{Lattice-based STIRAP transfer}
We implement two-photon STIRAP transfer to the deeply bound rovibrational level $|3\!> = |v\!=\!73, J\!=\!2\!>$ of the $^1\Sigma_g^+$ electronic ground state potential in a similar way as in our previous work \cite{Danzl2008a}, except that now the molecules are trapped at the individual wells of the optical lattice. In brief, laser $L_1$ near a wavelength of $1126$ nm, driving the transition from $|1\!> = |s\!>$ to $|2\!>$, where $|2\!>$ is a deeply bound level of the mixed $($A$^1\Sigma_u^+ - $b$^3\Pi_u) \ 0_u^+$ excited states, is pulsed on after laser $L_2$, which drives the transition from $|3\!>$ to $|2\!>$ at 1006 nm, see Fig.\,1 A. The pulse (or pulse overlap) time $\tau_p$ is typically $\tau_p = 10 \ \mu$s for the present experiments. A schematic time course for the transition Rabi frequencies is shown in Fig.\,3 C. We estimate the peak Rabi frequencies to be $2 \pi \times 3$ MHz for the transition at 1126 nm and $2 \pi \times 6$ MHz for the transition at 1006 nm \cite{Danzl2008a}. After a variable hold time $\tau_h$, we reverse the pulse sequence to transfer the molecules back to $|s\!>$. For short $\tau_h$ below $40 \ \mu$s we typically leave $L_1$ on between the two STIRAP pulse sequences. For longer $\tau_h$ we switch $L_1$ off to avoid any residual optical excitation of molecules in $|v\!=\!73, J\!=\!2\!>$ and possible effects of dipole forces generated by the tightly focused laser beam $L_1$.

The result of double STIRAP transfer in the optical lattice is shown in Fig.\,3 A. Here, $\tau_p \! = \! 10 \ \mu$s and $\tau_h \! = \! 15 \ \mu$s. As before, we interrupt the transfer after a given STIRAP time $\tau_S$ and record the number of molecules in the initial state $|s\!>$. The molecules first disappear, and then a sizable fraction of about 65\% returns after the reverse STIRAP transfer. Thus, as in our previous work \cite{Danzl2008a}, the single pass efficiency is about 80\% when both lasers are on resonance. Fig.\,3 B shows the double STIRAP transfer efficiency as a function of the detuning $\Delta_2$ of laser $L_2$ from the excited intermediate level while laser $L_1$ is held on resonance (detuning $\Delta_1 \approx 0$). A Gaussian fit yields a full width at half maximum of 830 kHz. With $\tau_p $ so short, we do not resolve molecular hyperfine structure in $|v\!=\!73, J\!=\!2\!>$.

We find that the molecules transferred to $|v\!=\!73, J\!=\!2\!>$ are trapped at the individual wells of the lattice. The $1/e$-lifetime is about $19$ ms for a lattice depth of 15 $E_R$. This is much shorter than the lifetime of Feshbach molecules as shown above, but sufficiently long to allow future implementation of a second lattice-based STIRAP step to the rovibronic ground state $|v\!=\!0, J\!=\!0\!>$, for which the lifetime is expected to be much longer as discussed below. We determine the lifetime by repeating the double STIRAP transfer while increasing the hold time $\tau_h$ in steps of 3.5 ms. The result is shown in Fig.\,2 D. The number of molecules can be well fit by an exponentially decaying function as a function of $\tau_h$. For a higher lattice depth of 25 $E_R$, the lifetime is reduced to $15$ ms. We thus attribute the reduced molecular lifetime to off-resonant scattering of lattice light, exciting the molecules to levels of the $($A$^1\Sigma_u^+ - $b$^3\Pi_u) \ 0_u^+$ states, which then in turn leads to loss into other ground state rovibrational levels that we do not detect.

\section{Determination of molecule trapping parameters}

We determine the molecular trapping frequency $\omega_{|v=73>}$ for molecules in $|v\!=\!73, J\!=\!2\!>$ by modulating the lattice phase and, alternatively, by modulating the lattice amplitude. In the first case, we primarily excite transitions from the lowest band in the lattice to the first excited band and then to higher bands. In the second case, we primarily excite into the second excited band and then to higher bands. For sufficiently strong modulation, molecules are lost from the lattice, as tunneling to neighboring sites and hence inelastic collisions with neighboring molecules become more probable. We thus expect to detect increased molecular loss if the modulation frequency is tuned into resonance with the inter-band transitions. The results are shown in Fig.\,4. At a lattice depth of 15 $E_R$, we observe resonant loss at 5.2 kHz in the case of phase modulation and at 10.1 kHz in the case of amplitude modulation of the lattice. Phase modulation at 22 $E_R$ and amplitude modulation at 20 $E_R$ yield resonances at 6.5 kHz and 12.2 kHz, respectively. These values for different trap depths are consistent with each other when compared with a calculation of the band structure. For comparison, to determine the trapping frequency $\omega_F$ of the Feshbach molecules in level $|g\!\!>$, we measure that phase modulation (amplitude modulation) of a 15 $E_R$ deep lattice leads to loss at a modulation frequency of 9.4 kHz (18.4 kHz). Relating the dynamical polarizability $\alpha_{|v=73>}$ of the deeply bound molecules in $v=73$ to the dynamical polarizability $\alpha_{F}$ of the Feshbach molecules via $\alpha_{|v=73>}/\alpha_F=\omega_{|v=73>}^2/\omega_F^2$, we obtain that the molecular polarizability in $|v\!=\!73, J\!=\!2\!>$ is $\sim 30$\% of the polarizability of the Feshbach molecules at the wavelength of our trapping light. Note that in the wavelength region of our trapping laser, this value is expected to show strong variations as a function of trapping laser wavelength due to the presence of levels of the $($A$^1\Sigma_u^+ - $b$^3\Pi_u) \ 0_u^+$ states.

\section{Conclusion}
We have transferred an ultracold sample of Cs$_2$ molecules to the deeply bound rovibrational level $|v\!=\!73, J\!=\!2\!>$ of the singlet X$^1\Sigma_g^+$ potential in the presence of an optical lattice. We essentially find the same transfer efficiency as in our previous work \cite{Danzl2008a} where no lattice was used. The transferred molecules are trapped, and we have determined their polarizability in this particular level. The trapping time is sufficiently long to allow for subsequent lattice-based STIRAP transfer to the rovibronic ground state $|v\!=\!0, J\!=\!0\!>$ by means of a second two-photon transition \cite{Mark2008}. A lower bound for the STIRAP pulse time and hence for the minimal required trapping time is set by the time needed to resolve the molecular hyperfine structure. This minimal time is the inverse of three times the ground state hyperfine coupling constant $c_4 \approx 14$ kHz \cite{Aldegunde2008}, giving $24 \ \mu$s. Hence a compromise can easily be found between Fourier-resolving the molecular hyperfine structure and keeping the STIRAP pulse time sufficiently short in view of finite laser coherence time and finite trapping time. For Cs$_2$ molecules in the rovibronic ground state $|v\!=\!0, J\!=\!0\!>$ we expect much longer trapping times in the lattice as optical excitation at $1064.5$ nm into excited molecular states can only occur in a far off-resonant process. At this wavelength transitions to the $($A$^1\Sigma_u^+ - $b$^3\Pi_u) \ 0_u^+$ states are relevant. These are possible only to levels that have a sizable singlet contribution stemming from the A$^1\Sigma_u^+$ state. $0_u^+$ levels below the minimum of the A$^1\Sigma_u^+$ state, corresponding to a wavelength of $\sim \! 1041$ nm as measured from the rovibronic ground state \cite{Verges1987}, have little singlet component and hence these transitions are strongly suppressed. We thus expect the formation of a stable molecular quantum gas in $|v\!=\!0, J\!=\!0\!>$ when the lattice depth is lowered and the molecules are released into a larger-volume optical dipole trap, possibly allowing the observation of Bose-Einstein condensation of ground state molecules.

\ack

We are indebted to R. Grimm for generous support and we thank S. Knoop, N. Boloufa, and O. Dulieu for valuable discussions. We gratefully acknowledge funding by the Austrian Ministry of Science and Research (BMWF) and the Austrian Science Fund (FWF) in form of a START prize grant. R.H. acknowledges support by the European Union in form of a Marie-Curie International Incoming Fellowship (IIF).

\clearpage

\begin{center}
\includegraphics[width=12cm]{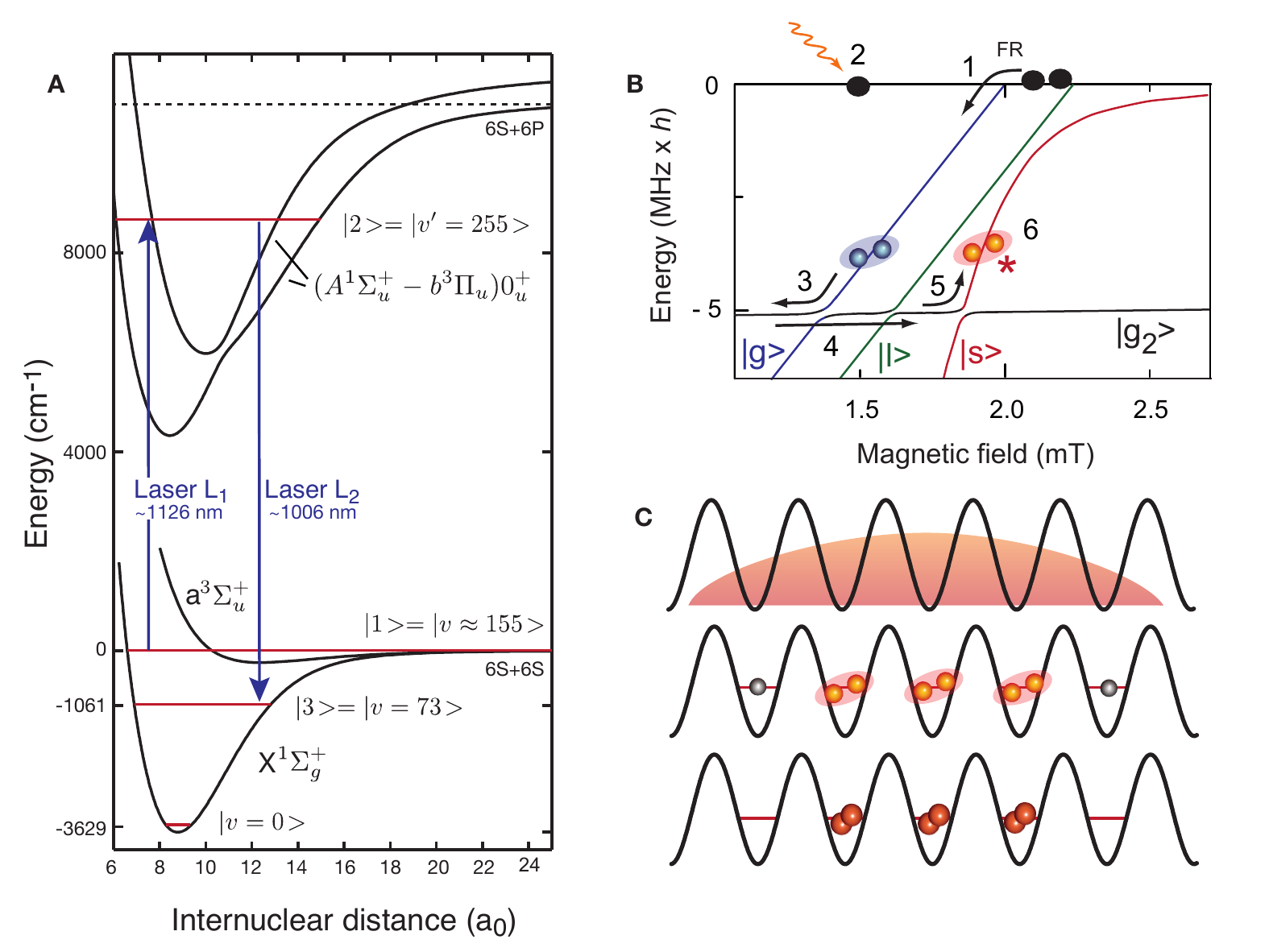}
\end{center}

\noindent {\bf Fig. 1.} {\bf A} Molecular level scheme for Cs$_2$. Molecules in a weakly bound Feshbach level $|1\!\!>$ are transferred to rovibrational level $|3\!\!>=|v\!=\!73,J\!=\!2\!\!>$ of the singlet X$^1\Sigma_g^+$ potential in the presence of an optical lattice. Level $|3\!\!>$ with a binding energy of $1061$ cm$^{-1}$ is reached in a two-photon STIRAP process with wavelengths near 1126 nm and 1006 nm via the 225th level of the electronically excited $ ($A$^1\Sigma_u^+ - $b$^3\Pi_u) \ 0_u^+$ potentials. The X$^1\Sigma_g^+$ potential has about $155$ vibrational levels. {\bf B} Zeeman diagram showing the energy of all relevant weakly bound molecular levels for initial Feshbach molecular state preparation \cite{Mark2007}. The binding energy is given with respect to the $F\!=\!3, m_F\!=\!3$ two-atom asymptote. The molecules are first produced on a g-wave Feshbach resonance at 1.98 mT in state $|g\!\!>$ (1). Residual atoms are removed by a combined microwave and resonant light pulse (2). The molecules are then transferred to the weakly bound s-wave state $|1\!\!>=|s\!\!>$ (6), the starting state for the STIRAP transfer, via three avoided state crossings involving state $|g_2\!\!>$ by slow (3,5) and fast magnetic field ramps (4). {\bf C} Lattice based ground state transfer. Top: The BEC is adiabatically loaded into the three-dimensional optical lattice, creating a Mott-insulator state. Middle: Atoms at doubly occupied sites are converted to Feshbach molecules. Atoms at singly occupied sites are removed thereafter. Bottom: The molecules are subsequently transferred to the deeply bound rovibrational level $|3\!\!>=|v\!=\!73,J\!=\!2\!\!>$ while shielded from collisions by the lattice potential.

\clearpage

\begin{center}
\includegraphics[width=12cm]{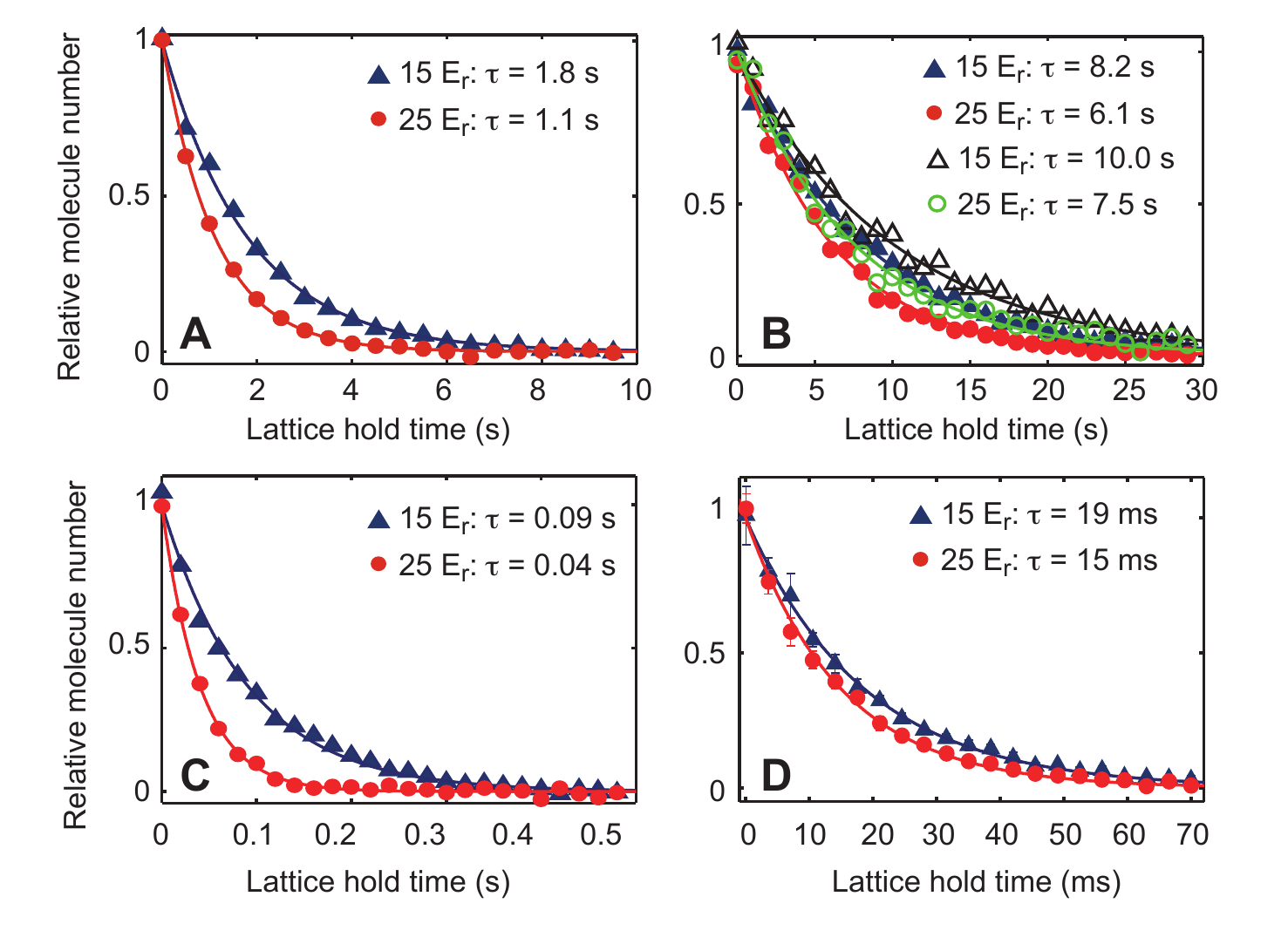}
\end{center}

\noindent {\bf Fig. 2.} Lifetime measurements of ultracold molecules trapped in the optical lattice. {\bf A}, {\bf B}, and {\bf C} show the decay of trapped Feshbach molecules, while {\bf D} shows the decay for molecules in $|3\!\!>=|v\!=\!73,J\!=\!2\!\!>$ of the X$^1\Sigma_g^+$ ground state potential.
In all cases, the triangles (circles) correspond to a lattice depth of 15 $E_R$ (25 $E_R$). All lifetimes $\tau$ are determined from exponential fits to the data as shown by the the solid lines. {\bf A} Lifetime of state $|g\!\!>$. {\bf B} Lifetime of state $|g_2\!\!>$ (filled symbols) and of state $|s\!\!>$ at $B\!=\!2.9$ mT (open symbols). {\bf C} Lifetime of state $|s\!\!>$ at $B\!=\!1.9$ mT, from where we drive the STIRAP transfer. {\bf D} Lifetime of molecules in the rovibrational level  $|3\!\!>=|v\!=\!73,J\!=\!2\!\!>$. The STIRAP lasers are switched off during the hold time in $|3\!\!>$. In {\bf D}, each data point is the average of 4 experimental runs, error bars correspond to the $1 \sigma$ statistical uncertainty. The typical uncertainty for the lifetimes is one unit of the last digit given.

\clearpage

\begin{center}
\includegraphics[width=12cm]{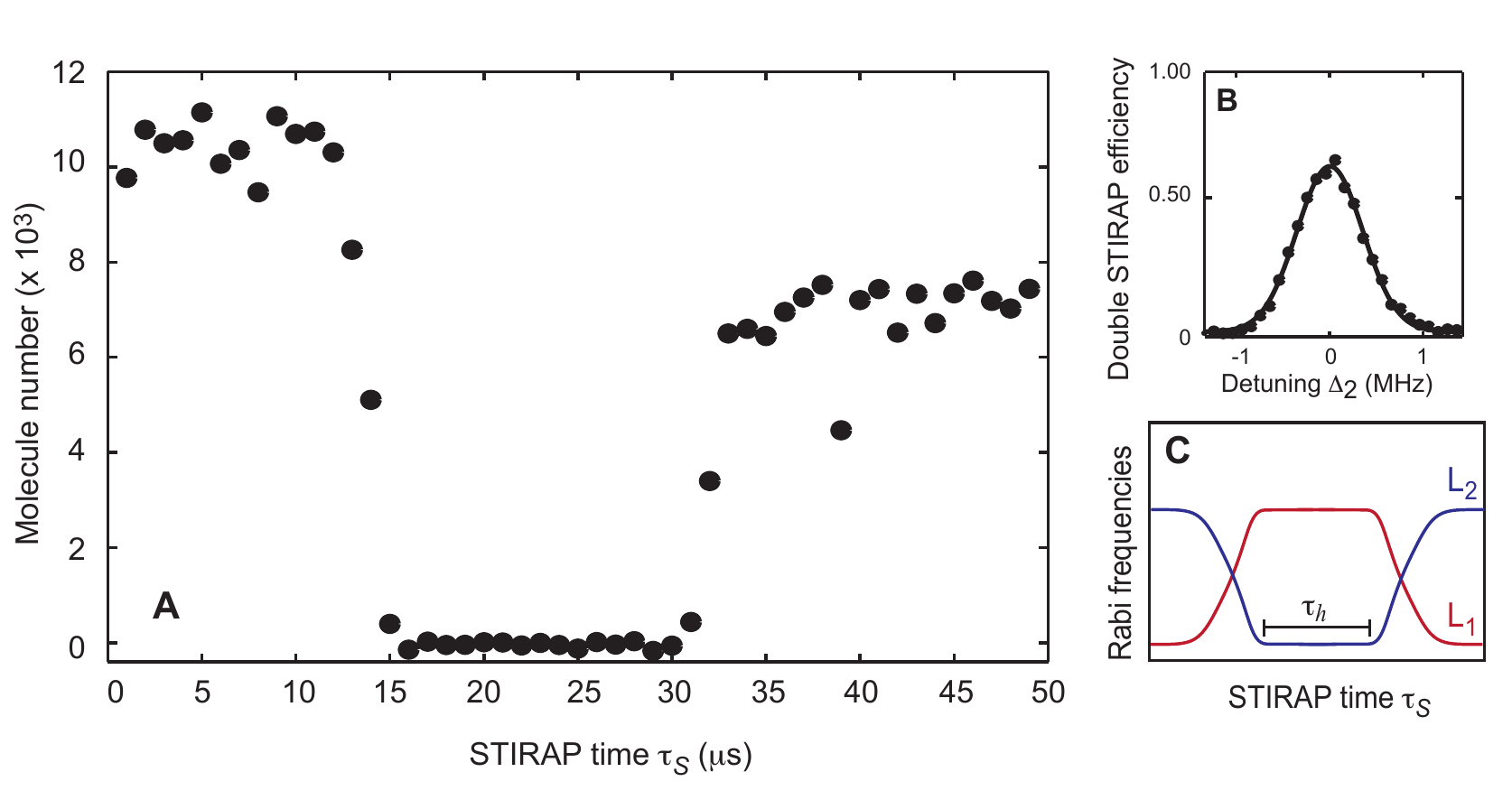}
\end{center}

\noindent {\bf Fig. 3.} STIRAP transfer from the weakly bound state $|1\!\!>=|s\!\!>$ to the deeply bound rovibrational level $|3\!\!>=|v\!=\!73, J\!=\!2\!\!>$ and back to $|1\!\!>$ in the optical lattice. {\bf A} shows the number of molecules in state $|1\!\!>$ as a function of STIRAP time $\tau_S$ for both lasers on resonance (laser detunings $\Delta_1 \approx 0 \approx \Delta_2$). The measured pulse overlap begins at about $5 \ \mu$s and ends at about $15 \ \mu$s. The second pulse overlap starts at $30 \ \mu$s and ends at about $38 \ \mu$s. The lattice depth is 15 $E_R$. Data points represent a single experimental realization, not an average over several runs. The data point at 39 $\mu$s represents a "bad shot", which occasionally occurs. {\bf B} Double STIRAP efficiency as a function of the detuning $\Delta_2$ of laser $L_2$ for $\Delta_1 \approx 0$. The solid line is a Gaussian fit with a full width at half maximum of $830$ kHz. {\bf C} schematically shows the timing for the Rabi frequencies, $\Omega_i$, $i=1,2$, for lasers $L_1$ and $L_2$ during the double STIRAP sequence. For short hold times $\tau_h < 40 \ \mu$s laser $L_1$ is left on after the first STIRAP sequence as shown here. For longer hold times $\tau_h > 40 \ \mu$s we shut off $L_1$ to avoid possible optical excitation.

\clearpage

\begin{center}
\includegraphics[width=12cm]{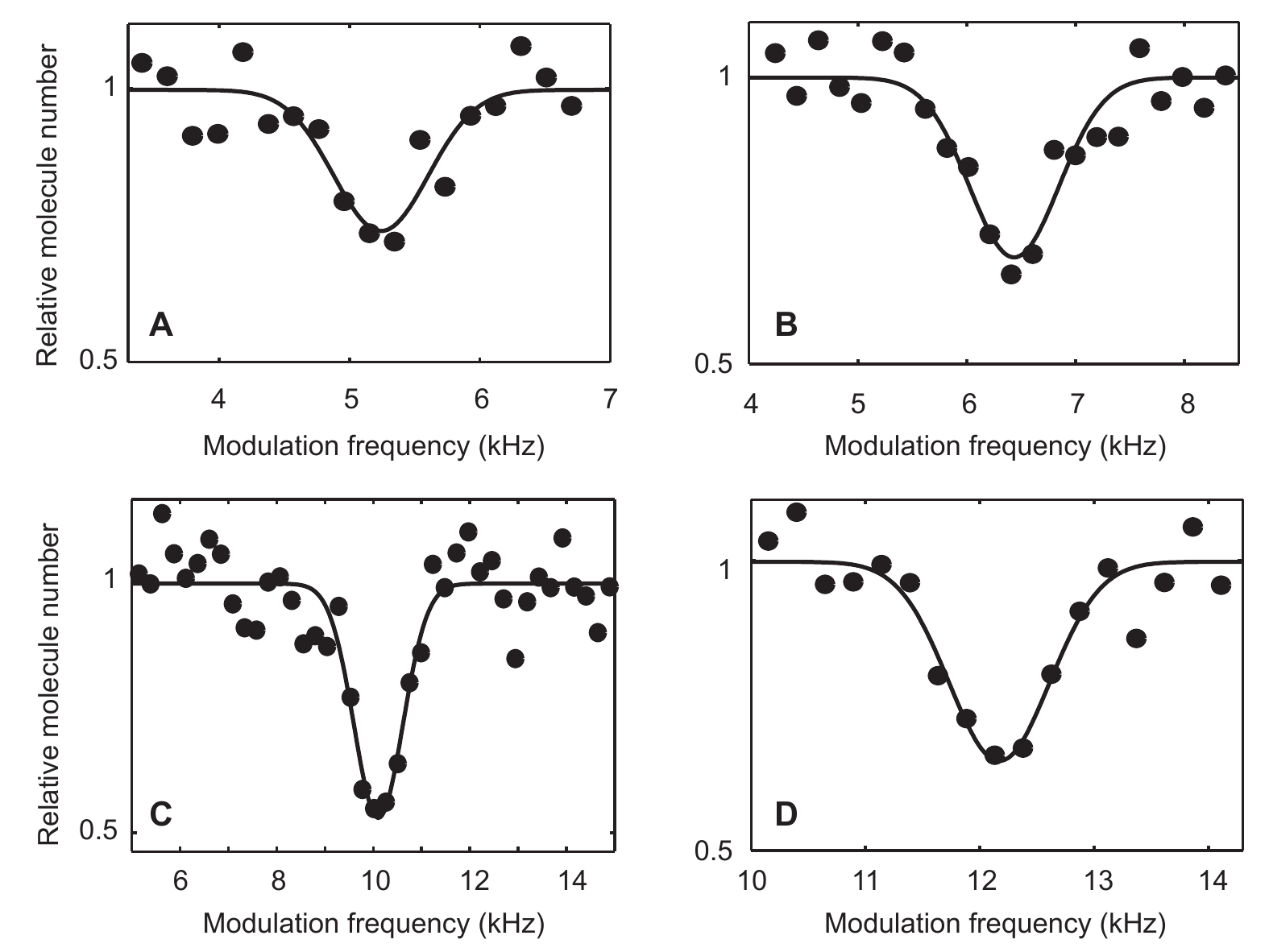}
\end{center}

\noindent {\bf Fig. 4.} Trapping of deeply bound molecules in the wells of the optical lattice. While the molecules reside in level $|3\!\!>=|v\!=\!73, J\!=\!2\!\!>$, one of the lattice beams of the 3 dimensional optical lattice is either phase modulated ({\bf A} and {\bf B}) or amplitude modulated ({\bf C} and {\bf D}). As the frequency of the phase or amplitude modulation is scanned, a series of resonances due to transfer to higher bands arise, reflected in a decrease in molecule number. The respective resonances at the lowest modulation frequency are shown here. For phase modulation ("shaking" of the lattice), this corresponds to the first lattice band, for amplitude modulation to the second band. To determine the center frequency, the resonances are fit by a Gaussian. The lattice depth is 15 $E_R$, 22 $E_R$, 15 $E_R$, and 20 $E_R$ in {\bf A}, {\bf B}, {\bf C}, and {\bf D}, respectively.

\end{document}